\documentclass[a4paper,11pt]{article}
\pdfoutput=1 % if your are submitting a pdflatex (i.e. if you have
\usepackage{jinstpub} % for details on the use of the package, please
\usepackage{siunitx}
\usepackage{lineno}
\title{CMS Phase-1 pixel detector refurbishment during LS2 and readiness towards the LHC Run~3}

\author[a,b]{Lars Olivier Sebastian Noehte$_{1}$\note{on behalf of the Tracker Group of the CMS collaboration}}
\affiliation[a]{Universität Zürich,\\Winterthurerstrasse 190, 8057 Zürich, Switzerland}
\affiliation[b]{Paul-Scherrer Institut,\\Forschungsstrasse 111, 5232 Villigen PSI, Switzerland}

\emailAdd{lars.noehte@cern.ch}

\abstract{
The CMS Phase-1 pixel detector was extracted from the underground cavern after the end of the LHC Run~2 in 2019 and has been kept cold to protect the silicon sensors during the long shutdown period (LS2) in 2019--2021. 
The LHC is now preparing for the next period of data taking, Run~3, which is scheduled to start in spring 2022. %for the next installment of the data taking beginning 2022. 
The Phase-1 pixel detector was going through a series of refurbishment and repairs this year to improve the quality of the collected data and enhance the detector performance. %operational experience. 
The innermost barrel pixel layer has been replaced with new modules and features improved front-end readout chips (PROC600v4), token bit manager chips (TBM10d), and circuit boards to rectify the issues discovered in Run~2. %during the previous data taking. 
The forward pixel detector has been equipped with new cooling inlets for safe handling and features a revised high-voltage power distribution scheme to better match the low-voltage granularity. 
All the DC-DC converter modules have been replaced with new modules featuring an improved FEAST2.3 ASIC, which is considerably more robust against the Total Ionizing Dose effect and thus prevents them from breaking during operation.
Overall, this article will summarize the refurbishment work of the pixel detector during LS2, and highlight the readiness towards the LHC Run~3 after installation and commissioning.
}

\keywords{Particle tracking detectors, Particle tracking detectors (Solid-state detectors), Large detector systems for particle and astroparticle physics}

\arxivnumber{2203.11525} % only if you have one

 \collaboration{\includegraphics[height=17mm]{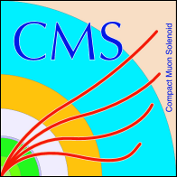}\\[6pt]}
\proceeding{PSD12: 12th International Conference on Position Sensitive Detectors\\
  12--17 Sep 2021\\
  University of Birmingham, Birmingham (United Kingdom)
  }

\begin{document}
\maketitle
\flushbottom

\section{Introduction}\label{sec:intro}
The CMS pixel detector is the innermost sub-detector of the CMS experiment~\cite{CMS}.
The pixel detector was upgraded in so-called Phase-1 upgrade, which took place during the long shutdown~1~(LS1) in 2016--2017, and it successfully took data over the period of 2017--2018 during the CERN Large Hadron Collider (LHC) Run~2~\cite{TDR_Phase_1}. 
At the time of writing, LHC is in the second long shutdown LS2 (2019--2021).
CMS uses this time for maintenance of the detector.
To extend its lifetime in the harsh radiation environment of the LHC, the LS2 was used to refurbish the Phase-1 pixel detector.

\subsection{CMS Phase-1 pixel detector}\label{sec:phase1}
The CMS Phase-1 pixel detector is closest to the collision point of the sub-detectors in CMS\@.
It provides excellent spatial particle track information for the CMS detector.
The Phase-1 pixel detector has over 124~million readout channels.
These channels are distributed over four barrel layers~(BPix), and three forward disks~(FPix) on each end~\cite{TDR_Phase_1}.
The pixel modules have a hybrid design with a silicon sensor connected to 16 front-end readout chips~(ROCs).
The outgoing data stream is routed on a high density interconnect~(HDI) and managed by an ASIC called token bit manager~(TBM).
The entire readout of a module in the innermost layer~(L1) is designed to cope with a particle hit rate of up to \SI{600}{\mega\hertz/\centi\metre\squared}.

\section{Refurbishments and testing}\label{sec:refurbishments}
\begin{figure}[htbp]
\centering % \begin{center}/\end{center} takes some additional vertical space
\includegraphics[width=.4\textwidth]{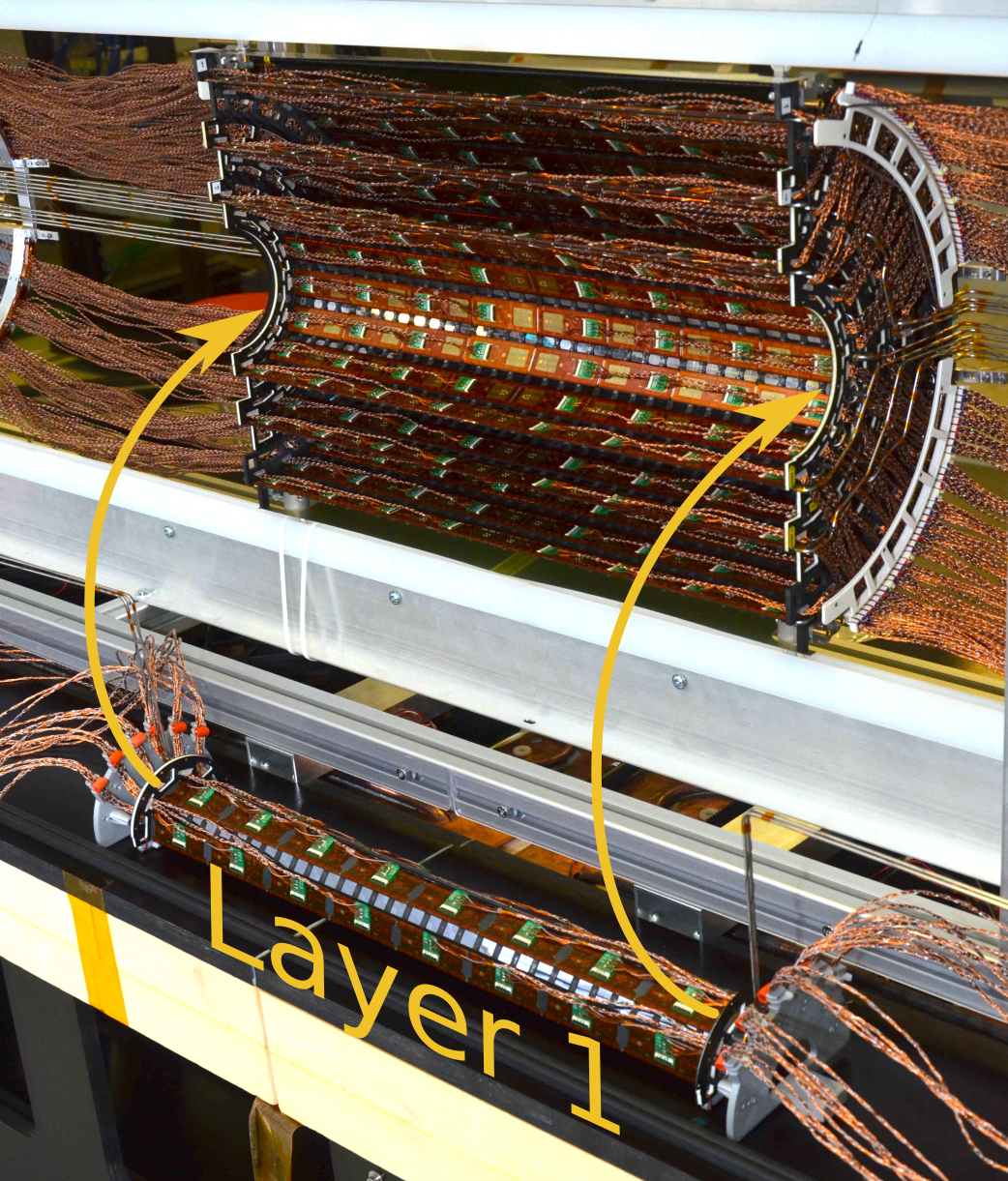}
\caption{\label{fig:Layer1install} This picture shows one half of BPix just before the first pixel layer was installed. L1 rests ready for installation in front of layers 2 through 4.}
\end{figure}

In BPix, L1 experiences the most radiation damage of all pixel layers.
During LHC Run~2 the Phase-1 pixel detector collected data for a recorded integrated luminosity of almost \SI{120}{\per\femto\barn}.
Expecting to take twice as much data during the upcoming Run~3 (2022--2025)~\cite{RUN3}, the fluence in the first pixel layer would exceed the operational limits.
The ROCs are designed to withstand a dose of 120\,Mrad, but have been tested beyond that~\cite{Phase_1_Upgrade}.
A new L1 was manufactured and installed in the BPix.
For this endeavor, the Phase-1 pixel detector was extracted from CMS in 2019 and stored in a cold and dry environment in a clean room.
The new installed modules feature improved ROCs called PROC600v4~\cite{Phase_1_Upgrade} and TBM ASICs called TBM10d~\cite{Phase_1_Upgrade}, rectifying issues discovered during Run~2.
The improved ROCs have better shielding of the calibration pulse injection circuit leading to reduced pixel cross-talk and noise.
Moreover, they have an improved time-stamp buffer logic, which in the past caused timing errors and thus loss of data synchronization.
The old TBM suffered from a vulnerability for a specific single event upset (SEU) which required a power cycle of the TBM to recover.
The new TBM is guarded against this SEU phenomena.
In addition, L1 is equipped with new high voltage cables with better insulation.
With the new cables and upgraded power supplies, the reverse bias voltage can be ramped up to over \SI{800}{\volt}, compared to \SI{450}{\volt} in Run~2.
A greater range for the high voltage in combination with a lower noise level helps to operate the modules at a very high efficiency~\cite{Irradiation} through Run~3.
Prior to the L1 installation, the inner modules of the second layer (L2) were easily accessible.
Thus, defect inner L2 modules were replaced.
Figure~\ref{fig:Layer1install} shows one half of BPix prepared for the installation of the new first pixel layer.
\newline
Some of the CO$_2$ cooling pipes of the FPix were noted to be broken at the connection to the end flanges.
Thus, all inlet connections were replaced with a more robust connection system.
For FPix it was also possible to harmonize the low and high voltage supply lines.
\newline
For both BPix and FPix, the controller ASIC on the DC-DC converter boards suffered during Run~2 from radiation induced leakage current in a transistor.
This caused charge build-up, damaging the connected chips.
All DC-DC converters were replaced with new DC-DC converters featuring an improved and more radiation hard ASIC called FEAST2.3~\cite{Phase_1_Upgrade}.
Concurrently everything was examined for faulty connections and fixed if needed.
A more detailed description of the improvements of the Phase-1 pixel detector can be found in Ref.~\cite{Phase_1_Upgrade}.
\newline
All modules were tested after reassembling the BPix in the clean room.
The testing procedure involved checking all power connections as well as electrical and optical data streams.
After synchronization of the configuration signals with the ROCs was achieved, the ROCs were tested up to a functioning pixel level.
These tests were performed first at room temperature and later at around \SI{-20}{\celsius}.
As a result of refurbishing, and testing, and resolving issues in the clean room, the fraction of working pixels is again higher than 99.1\,\% for BPix and 98.5\,\% for FPix, compared to 93.5\,\% for BPix and 96.7\,\% for FPix at the end of Run~2.

\section{Installation and checkout}\label{sec:installation}
After BPix and FPix were tested in the clean room and ready to be installed, BPix and FPix were moved to the cavern of the CMS experiment located \SI{100}{\metre} underground.
By design BPix had to be installed as two half barrels, since the beam pipe was already installed at the center of the experiment.
The two halves were guided on rails to carefully slide in place without touching the beam pipe or other components of the detector.
The reinsertion went without major difficulties and BPix was centered around the beam pipe.
All power and high voltage cables, as well as all optical fibers, were connected and BPix was made ready for testing.
\newline
After BPix was completely installed a quick checkout was performed.
Faulty power supplies had to be exchanged and optical fibers were examined, and cleaned, and non-functional ones replaced with spares.
Module testing after facilitating all the connections between BPix and the service cavern showed that the detector did not have any faults due to the moving and installation.
After the first checkout of BPix, FPix was installed and commissioned.
After commissioning at room temperature was finished, the detector volume was sealed and cooled down to about \SI{-20}{\celsius}.
After extensive testing up to the functioning pixel level, the fraction of working pixels stayed to be larger than 99.1\,\% for BPix and 98.5\,\% for FPix.
Soon after the first calibrations at cold temperature, large amount of data with over 3.5 million cosmic particle tracks was recorded with the refurbished detector.
\newline
In the second half of October 2021, the pixel detector successfully measured the first collision data at the LHC injection energy after extensive calibration of thresholds and other pixel operation parameters.

%\acknowledgments
%
%This is the most common positions for acknowledgments. A macro is
%available to maintain the same layout and spelling of the heading.
%
%\paragraph{Note added.} This is also a good position for notes added
%after the paper has been written.


\end{document}